\documentclass[pra,aps,twocolumn,showpacs]{revtex4}
\usepackage{graphicx}

\begin{document}

\title{Generation of maximum spin entanglement
induced by a cavity field\\ in quantum-dot systems}

\author{Adam Miranowicz,$^{1,2,3}$ \c{S}ahin K. \"Ozdemir,$^{1,2}$
Yu-xi Liu$^{2}$, Masato Koashi,$^{1,2}$ Nobuyuki
Imoto,$^{1,2,4,5}$\\ and Yoshiro Hirayama$^{1,5}$}

\address{
$^1$CREST Research Team for Interacting Carrier Electronics,
Japan Science and Technology Corporation, Tokyo, Japan\\
$^2$School of Advanced Sciences, Graduate University for Advanced
Studies (SOKENDAI), Hayama, Kanagawa 240-0193, Japan\\
$^3$Nonlinear Optics Division, Institute of Physics, Adam
Mickiewicz University, 61-614 Pozna\'n, Poland\\
$^4$University of Tokyo, 7-3-1 Hongo, Bunkyo-ku, Tokyo 113-8654,
Japan\\
$^5$NTT Basic Research Laboratories, 3-1 Morinosato Wakamiya,
Atsugi, Kanagawa 243-0198, Japan}

\date{19 June 2002. Published in: {\em Physical Review A} {\bf 65} (2002) 062321}
\pagestyle{plain} \pagenumbering{arabic}

\begin{abstract}

Equivalent-neighbor interactions of the conduction-band electron
spins of quantum dots in the model of Imamo\={g}lu {\em et al.}
[Phys. Rev. Lett. {\bf 83}, 4204 (1999)] are analyzed. An
analytical solution and its Schmidt decomposition are found and
applied to evaluate how much the initially excited dots can be
entangled with the remaining dots if all of them are initially
disentangled. It is demonstrated that perfect maximally entangled
states (MESs) can only be generated in systems of up to six dots
with a single dot initially excited. It is also shown that highly
entangled states, approximating the MESs with good accuracy, can
still be generated in systems of odd numbers of dots with almost
half of them excited. A sudden decrease of entanglement is
observed on increasing the total number of dots in a system with a
fixed number of excitations.
\end{abstract}

\pacs{03.65.Ud, 03.67.-a, 68.65.Hb, 75.10.Jm}

\maketitle

\pagenumbering{arabic}

\section{Introduction}

Since the seminal papers of Obermayer, Teich, and Mahler
\cite{Obe88}, there has been growing interest in the
quantum-information properties of quantum dots (QDs) in the quest
to implement quantum-dot scalable quantum computers
\cite{QD-QC,Loss98,QD-spin}. Those high expectations are justified
to some extent by recent experimental advances in the coherent
observation and manipulation of quantum dots \cite{Bon98,Oos98},
including spectacular demonstrations of the quantum entanglement
of excitons in a single dot \cite{Chen00} or quantum-dot molecule
\cite{Bay01}, and observations of Rabi oscillations of excitons in
single dots \cite{Kam01}. Among various models of quantum
computers based on localized electron spins of quantum dots as
qubits \cite{Loss98,QD-spin}, the scheme of Imamo\={g}lu {\em et
al.} \cite{Ima99} is the first, where the interactions between the
qubits are mediated by a cavity field. This  approach combines the
advantages of long-distance optically controlled couplings with
long-decoherence times of the spin degrees of freedom. Here, we
analyze quantum entanglement in the Imamo\={g}lu {\em et al.}
model.

During the last decade, it has been highlighted that quantum
entanglement, being at the heart of quantum mechanics, is also a
powerful resource for quantum communication and
quantum-information processing. Quantum entanglement in
interacting systems is a common phenomenon. It is obvious that any
interacting many-body system with defined qubits, if set in a
properly chosen state, will evolve through states with entangled
qubits. Surprisingly, quantitative descriptions of the
entanglement dynamics in multiparticle systems are by no means
satisfactory yet \cite{multipart}. Nevertheless, in a special case
of bipartite entanglement, a number of measures have been
introduced and studied \cite{Ben96,Ved97,Woo98}. For example,
entanglement of a bipartite system in a pure state, described by
the density matrix $\hat{\rho}_{AB}=(|\psi \rangle \langle \psi
|)_{AB}$, can be measured by the von Neumann entropy
\cite{Ben96,Ved97}
\begin{equation}
E[\hat{\rho}_{AB}]=-{\rm Tr}\{\hat{\rho}_{A}\log
_{2}\hat{\rho}_{A}\}=-{\rm Tr}\{\hat{\rho}_{B}\log
_{2}\hat{\rho}_{B}\} \label{N01}
\end{equation}
of the reduced density matrix $\hat{\rho}_{A}={\rm
Tr}_{B}\{\hat{\rho}_{AB}\}$ or, equivalently, $\hat{\rho}_{B}={\rm
Tr}_{A}\{\hat{\rho}_{AB}\}$. The entanglement of formation of a
mixed state of a bipartite system is often measured by the
so-called concurrence proposed by Hill and Wootters\cite{Woo98}.
Concurrence has been applied to study entanglement in various
models \cite{conc} including equivalent-neighbor systems
\cite{Koa00,Wang01}. The following two aspects of entanglement are
especially important: (i) coherent manipulation of entanglement
and (ii) generation of maximum entanglement. The possibility of
coherent and selective control of entanglement in a quantum-dot
system was analyzed by Imamo\={g}lu {\em et al.} \cite{Ima99}.
Here, we would like to focus on the latter topic, i.e., the
generation of the maximally entangled states (MESs) of quantum
dots in the model of Imamo\={g}lu {\em et al.} \cite{Ima99}. MESs
are necessary for the majority of quantum information-processing
applications. Otherwise, for example, direct application of partly
entangled states for teleportation will result in unfaithful
transmission, while superdense coding with partly entangled states
will cause noise in the resulting classical channel.

The paper is organized as follows. In Sec. II, we describe an
equivalent-neighbor quantum-dot model and give its analytical
solution. In Sec. III, we analyze the possibilities of generation
of the MESs or their good approximations for different initial
conditions of the number of excitations and the total number of
dots in the system.

\vspace*{0mm}
\begin{figure}
\includegraphics[width=7cm]{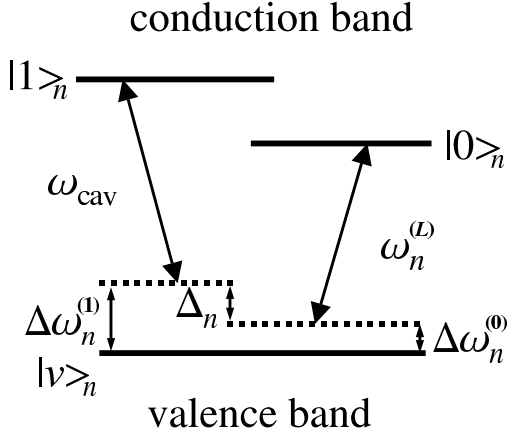}
\caption{Three-level atom in $V$ configuration as a model of a
semiconductor dot with the conduction-band spin states $|1\rangle
_{n}$ (spin up) of energy ${\cal E} _{n}^{(1)}$ and $|0\rangle
_{n}$ (spin down) of energy ${\cal E} _{n}^{(0)}$ , and the
effective valence-band state $|v\rangle _{n}$ of energy ${\cal E}
_{n}^{(v)}$ in the $n$th dot. Key: $\omega _{{\rm cav}}$,
frequency of the common cavity mode; $\omega _{n}^{(L)}$,
frequency of the classical laser field addressed at the $n$th dot;
$\hbar\Delta \omega _{n}^{(1)}={\cal E}_{n}^{(1)}-{\cal E}
_{n}^{(v)}-\hbar\omega _{{\rm cav}}$, $\hbar\Delta \omega
_{n}^{(0)}={\cal E} _{n}^{(0)}-{\cal E} _{n}^{(v)}-\hbar\omega
_{n}^{(L)}$, and $\Delta _{n}=\Delta \omega _{n}^{(1)}-\Delta
\omega _{n}^{(0)}$ are detunings. }
\end{figure}
\section{Quantum-dot model and its solution}

We will apply the model of Imamo\={g}lu {\em et al.} \cite{Ima99}
to describe strong equivalent-neighbor couplings of quantum-dot
spins through a single-mode microcavity field. The dots are placed
inside a microdisk, put into a microcavity tuned to frequency
$\omega _{{\rm cav}}$, and illuminated selectively by laser fields
of frequencies $\omega _{n}^{(L)}$. Each of $N$ dots with a single
electron in the conduction band is modeled by a three-level atom
as shown in Fig. 1. The total Hamiltonian for $N$ three-level
quantum dots interacting with $N+1$ quantized fields reads
\begin{eqnarray} \label{N02}
\hat{H}&=&\hat{H}_{QD}+\hat{H}_{F}+\hat{H}_{\rm int},
\\
\hat{H}_{QD}&=&\sum_{n}({\cal E}_{n}^{(0)}\hat{\sigma}
_{n}^{00}+{\cal E}_{n}^{(1)}\hat{\sigma} _{n}^{11}+{\cal
E}_{n}^{(v)}\hat{\sigma} _{n}^{vv}),
  \nonumber \\
\hat{H}_{F}&=&\hbar \omega _{{\rm cav}}\hat{a}_{{\rm cav}}^{\dag
}\hat{a}_{{\rm cav} }+\sum_{n}\hbar \omega
_{n}^{(L)}(\hat{a}_{n}^{(L)})^{\dag }\hat{a}_{n}^{(L)},
  \nonumber \\
\hat{H}_{\rm int} &=&\sum_{n}\hbar
g_{n}^{v0}[\hat{a}_{n}^{(L)}\hat{\sigma}
_{n}^{0v}+(\hat{a}_{n}^{(L)})^{\dag }\hat{\sigma} _{n}^{v0}]
  \nonumber \\
&&+\sum_{n}\hbar g_{n}^{v1}(\hat{a}_{{\rm cav}}\hat{\sigma}
_{n}^{1v}+\hat{a}_{{\rm cav} }^{\dag }\hat{\sigma} _{n}^{v1}),
  \nonumber
\end{eqnarray}
where $\hat{H}_{QD}$ and $\hat{H}_{F}$ are the free Hamiltonians
of the quantum dots and the fields, respectively; $\hat{H}_{\rm
int}$ is the interaction Hamiltonian; $\hat{a}_{\rm cav}$ and
$\hat{a}^{\dag}_{\rm cav}$ are the annihilation and creation
operators of the cavity mode, respectively; $\hat{a}^{(L)}_{n}$
and $(\hat{a}^{(L)}_{n})^{\dag}$ are the corresponding operators
for the laser modes; $\hat{\sigma} _{n}^{xy}$ is the $n$th dot
operator given by $\hat{\sigma} _{n}^{xy}=|x\rangle _{nn}\langle
y|$; ${\cal E}_{n}^{(x)}$ is the energy of level $|x\rangle_{n}$
($x=0,1,v$); the $n$th dot levels $|0\rangle _{n}$ and $|v\rangle
_{n}$ are coupled by dipole interactions with a strength of
$g_{n}^{v0}$; analogously, $g_{n}^{v1}$ is the coupling strength
between levels $|1\rangle _{n}$ and $ |v\rangle _{n}$. There is no
direct coupling between levels $|0\rangle _{n} $ and $|1\rangle
_{m}$ in either the same ($n=m$) or different dots ($n\neq m$).
The Hamiltonian (\ref{N02}) simply generalizes, to $N$ dots and
$N+1$ fields, models of a three-level atom (dot) interacting with
two modes of radiation fields widely discussed in the literature
(see, e.g., \cite{3plus2}). By applying an adiabatic elimination
method, Imamo\={g}lu {\em et al.} derived the effective
interaction Hamiltonian describing the evolution of the
conduction-band spins of $N$ quantum dots coupled by a microcavity
field in the form \cite{Ima99}
\begin{eqnarray}
\hat{H}_{{\rm eff}}&=&\!\frac{\hbar}{2}\sum_{n\neq m}\kappa
_{nm}(t)[{\hat{\sigma}} _{n}^{+}{\hat{\sigma}}_{m}^{-}e^{i(\Delta
_{n}-\Delta _{m})t} \nonumber \\ &&~~~~~~~~~~~~~~~~+{\hat{\sigma}}
_{n}^{-}{\hat{\sigma}}_{m}^{+}e^{-i(\Delta _{n}-\Delta _{m})t}]
\label{N02a}
\end{eqnarray}
in terms of the Pauli spin creation ${\hat{\sigma}}_{n}^{+}$ and
annihilation ${\hat{\sigma}}_{n}^{-}$  operators acting on the
conduction-band spin states of the $n$th dot. The effective
two-dot coupling strength between the spins of the $n$th and $m$th
dots is given by $\kappa _{nm}(t)= g_n(t) g_m(t)/\Delta _{n}$,
where the effective single-dot coupling of the $n$th spin to the
cavity field is $g_n(t)= g_{n}^{v0} g_{n}^{v1}
|E^{(L)}_n(t)|/\Delta \omega _{n}$ with $\Delta \omega _{n}$ being
the harmonic mean of $\Delta \omega _{n}^{(1)}$ and $\Delta \omega
_{n}^{(0)}$. For simplicity, the laser fields are assumed to be
strong and treated classically as described by the complex
amplitudes $E^{(L)}_n(t)$. The Hamiltonian (\ref{N02a}) was
derived by applying adiabatic eliminations of the valence-band
states $|v\rangle _{n}$ and cavity mode $\hat{a}_{{\rm cav}},$
which are valid under the assumptions of negligible coupling
strength, cavity decay rate, and thermal fluctuations in
comparison to $\hbar\Delta _{n}$ and $\hbar\Delta \omega
_{n}^{(x)}$ ($x=0,1$) and the energy difference ${\cal
E}_{n}^{(1)}-{\cal E} _{n}^{(0)}$  (see Fig. 1). Moreover, the
valence-band levels $|v\rangle _{n} $ were assumed to be far off
resonance. Although the Hamiltonian (\ref{N02a}) describes
apparently direct spin-spin interactions, the real physical
picture is different: Quantum-dot spins are coupled only
indirectly via the cavity and laser fields.

Imamo\={g}lu {\em et al.} \cite{Ima99} applied their model  for
quantum computing purposes by implementing the conditional
phase-flip and controlled-NOT (CNOT) operations between two
arbitrary dots addressed selectively by laser fields to satisfy
the condition $\Delta _{n}=\Delta _{m}$. Here, we are interested
in a realization of an equivalent-neighbor model scalable for a
large number of dots (even for more than 100 \cite{Ima99}). This
goal can readily be achieved by assuming that all dots are
identical and illuminated by a single-mode stationary laser field
of frequency $\omega _{n}^{(L)}\equiv \omega ^{(L)}$, which
implies $\kappa _{nm}(t)=\kappa=\,$const. In fact, the condition
of equivalent-neighbor interactions can also be assured for
nonidentical dots by adjusting the laser-field frequencies $\omega
_{n}^{(L)}$ to get the same detuning $\Delta _{n}=\,$const, and by
choosing the proper laser intensities $|E^{(L)}_n|^2$ to obtain
the effective coupling constants of $g_{n}(t)=\,$const or,
equivalently, $\kappa _{nm}(t)=\,$const for every pair of dots.
Thus, Eq. (\ref{N02a}) can be reduced to the effective
equivalent-neighbor $N$-dot Hamiltonian as
\vspace*{0mm}
\begin{figure}
\includegraphics[width=8cm]{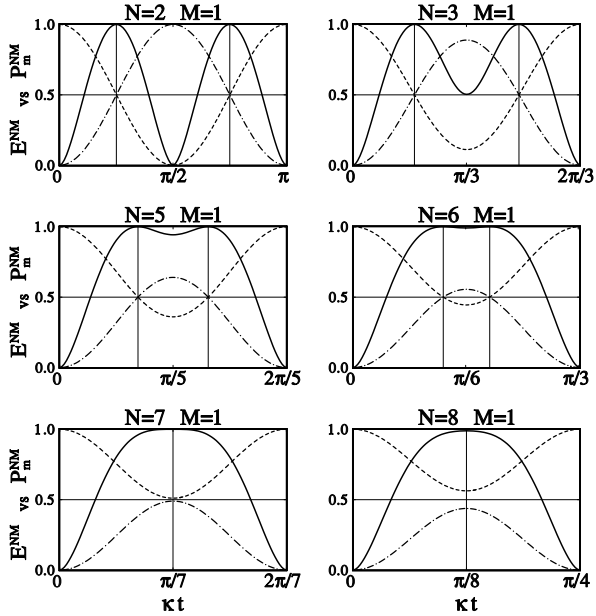}
\caption{Evolution of the quantum entanglement of $E^{N1}(t)$
(solid) and the Schmidt coefficients of $P_{0}^{N1}(t)$ (dashed)
and $P_{1}^{N1}(t)$  (dot-dashed curves) for systems of
$N=2,\cdots,8$ quantum dots with only one ($M=1$) of them
initially excited. Figure illustrates that the exact maximally
entangled states can be generated in systems of $N$ up to 6 dots
only.}
\end{figure}
\begin{eqnarray}
\hat{H}_{\rm eff}=\frac{\hbar\kappa}{2} \sum_{n\neq m}\left(
{\hat{\sigma}}_n^{+}{\hat{\sigma}}_m^{-}+
{\hat{\sigma}}_n^{-}{\hat{\sigma}}_m^{+}\right) \label{N03}
\end{eqnarray}
where $\kappa$ is the coupling constant. The system described by
Eq. (\ref{N03}) is sometimes referred to as the spin-$1/2$ van der
Waals model~\cite{Dek79}, the infinitely coordinated
system~\cite{Bot82}, the Lipkin or Lipkin-Meshkov-Glick
model~\cite{Lip65}, or just the equivalent-neighbor model
\cite{Liu90}.  Let us assume that the initial state describing a
system of $M$ ($M=0,\cdots ,N$) dots initially excited (i.e., with
conduction-band spins up) and $N-M$ dots in the ground state
(conduction-band spins down) is given as
\begin{eqnarray}
|\psi (0)\rangle  &=&\{|1\rangle ^{\otimes M}\}_{A}\{|0\rangle
^{\otimes (N-M)}\}_{B} \nonumber \\ &\equiv& |\underbrace{11\cdots
1}_{M}\rangle _{A}|\underbrace{00\cdots 0}_{N-M}\rangle _{B}.
\label{N04}
\end{eqnarray}%
Then, we find the solution of the Schr\"odinger equation of motion
for the model (\ref{N03}) in the form
\begin{eqnarray}
|\psi (t)\rangle &=&\sum_{m=0}^{M^{\prime }}C_{m}^{NM}(t)
{\{|1\rangle ^{\otimes (M-m)}|0\rangle ^{\otimes m}\}}_{A}  \nonumber \\
&&~~~~~~~~~~~~\otimes {\{|1\rangle ^{\otimes m}|0\rangle ^{\otimes
(N-M-m)}\}}_{B}, \label{N05}
\end{eqnarray}%
where $M^{\prime }=\min (M,N-M)$. The states in curly brackets
$\{|1\rangle ^{\otimes (n-m)}|0\rangle ^{\otimes m}\}$ denote the
sum of all $n$-dot states with $(n-m)$ excitations. For example,
${ \{|1\rangle ^{\otimes 2}|0\rangle ^{\otimes 2}\}}$ stands for
$|0011\rangle +|0101\rangle +|0110\rangle +|1001\rangle
+|1010\rangle +|1100\rangle$. The number of states in the
superposition $\{|1\rangle ^{\otimes (n-m)}|0\rangle ^{\otimes
m}\}$ (or equivalently $\{|1\rangle ^{\otimes m}|0\rangle
^{\otimes (n-m)}\}$) is equal to the binomial coefficient
$\textstyle{n \choose m}$. Thus, for given $N$ and $M$, the
solution (\ref{N05}) contains $\textstyle{N \choose M}$ terms. The
energy of the QD system described by Eq. (\ref{N03}) is conserved;
thus all the superposition states in Eq. (\ref{N05}) have the same
number $M$ of excitations. We find the time-dependent
superposition coefficients in Eq. (\ref{N05}) as
\begin{eqnarray}
C_{m}^{NM}(t)&=&\sum_{n=0}^{M^{\prime }}b_{nm}^{NM}\exp
\big\{i[n(N+1-n)
\nonumber \\
&&~~~~~~~~~~~~~~~~-M(N-M)]\kappa t\big\}  \label{N06}
\end{eqnarray}
in terms of
\begin{eqnarray}
b_{nm}^{NM}&=&\sum_{k=0}^{m} (-1)^{k} {m \choose k} {N-2k
\choose M-k}^{-1}  \nonumber \\
&&~~~\times \left[{N+1-2k \choose n-k}-2{N-2k \choose n-k-1}
\right] , \label{N07}
\end{eqnarray}
where $\textstyle{x \choose y}$ are binomial coefficients. Our
solution can be represented in a biorthogonal form via the Schmidt
decomposition
\begin{equation}
|\psi (t)\rangle =\sum_{m=0}^{M^{\prime
}}\sqrt{P_{m}^{NM}(t)}|\phi _{m}(t)\rangle _{A}\otimes |\varphi
_{m}(t)\rangle _{B},  \label{N08}
\end{equation}
where $|\phi _{m}(t)\rangle _{A}$ and $|\varphi _{m}(t)\rangle
_{B}$ are the orthonormal basis states of subsystems $A$ and $B$,
respectively. We find that the real and positive Schmidt
coefficients can be related to the squared module of superposition
coefficients (\ref{N06}) as follows:
\vspace*{0mm}
\begin{figure}
\includegraphics[width=8cm]{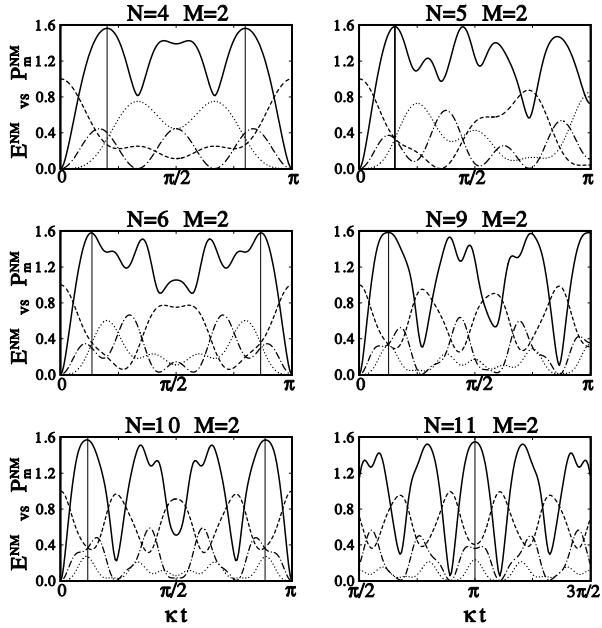}
\caption{Evolution of the entanglement of $E^{N2}(t)$ (solid) and
all Schmidt coefficients $P_{0}^{N2}(t)$  (dashed),
$P_{1}^{N2}(t)$  (dot-dashed), and $P_{2}^{N2}(t)$  (dotted
curves), in systems with two  ($M=2$) dots initially excited.}
\end{figure}
\begin{equation}
P_{m}^{NM}(t)={M \choose m} {N-M \choose m} |C_{m}^{NM}(t)|^{2},
\label{N09}
\end{equation}
while the phases of $C_{m}^{NM}(t)$ are absorbed into the
definition of the basis states  $|\phi _{m}(t)\rangle _{A}$ and
$|\varphi _{m}(t)\rangle _{B}$. The Schmidt coefficients are
normalized to unity. The evolutions of all $P_{m}^{NM}$ for
systems with single and two excitations are given in Figs. 2 and
3, respectively. We observe that the evolution of Schmidt
coefficients is periodic with the period of $\kappa T=2\pi/N$ for
systems with a single ($M=1$ or, equivalently, $M=N-1$) excitation
(Fig. 2), and $\pi$ periodical ($2\pi$ periodical) for systems of
even (odd) numbers of dots with higher numbers of excitations (see
Fig. 3). For brevity, only half of the period is depicted in the
right-hand panels of Fig. 3.

\section{Entanglement in quantum-dot systems}

We address the following questions: How much can the initially
excited dots (say, subsystem $A$) be entangled with the remaining
dots (subsystem $B$) in the equivalent-neighbor system of
initially all disentangled dots if the evolution is governed by
Hamiltonian (\ref{N03})? And whether the maximally entangled
states can be generated exactly or, at least, approximately in
systems of an arbitrary number $N$ of dots while $M$ of them are
excited.

With the help of an explicit form of the Schmidt decomposition, it
is convenient to calculate the  entanglement (\ref{N01}) via the
Shannon entropy
\begin{eqnarray}\label{N10}
E^{NM}(t)&\equiv& E[|\psi (t)\rangle \langle \psi (t)|] \nonumber \\
&=&-\sum_{m=0}^{M'}P_{m}^{NM}(t)\log _{2}P_{m}^{NM}(t)
\end{eqnarray}
of the Schmidt coefficients given for our system by (\ref{N09}).
By applying Eq. (\ref{N10}), we can determine the maximum
entanglement given by $E^{NM}_{\max}(t)\equiv \max_t E^{NM}(t)$,
which can periodically be generated during the evolution of
$N$-dot system with $M$ excitations. The coefficients (\ref{N09}),
as well as (\ref{N06}), possess the symmetry of
$P_{m}^{NM}(t)=P_{m}^{N,N-M}(t)$, which implies equal evolutions
of entanglement
\begin{equation}
E^{NM}(t)=E^{N,N-M}(t) \label{N11}
\end{equation}
in the $N$-dot systems with $M$ and $N-M$ excitations. Figure 4
shows this symmetry in a special case for maximum entanglement of
$\max_t E^{NM}(t)=\max_t E^{N,N-M}(t)$.
\vspace*{0mm}
\begin{figure}
\includegraphics[width=7cm]{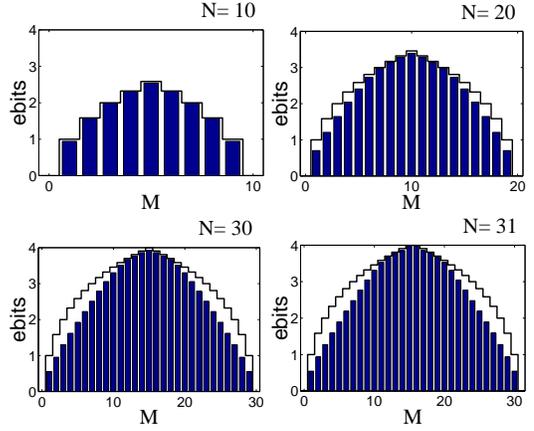}
\caption{Maximum entanglement $E_{\max }^{NM}=\max_{t}E^{NM}(t)$
(solid bars), measured in ebits,  as a function of the excitation
number $M$ generated in systems of $N=10,20,30$ and 31 dots. The
empty staircase corresponds to entanglement of $E_{\rm MES}^{NM}$
for the MESs. The figure illustrates that the highest
entanglement, closest to $E_{\rm MES}^{NM}$, can be generated in
systems with $M=[N/2]$ excitations. On decreasing $M$ or ($N-M$),
the entanglement decreases. The discrepancy between $E_{\max
}^{NM}$ and $E_{\rm MES}^{NM}$  becomes more pronounced with
increasing $N$ especially for $0<M\ll [N/2]$.}
\end{figure}

To solve the second problem proposed at the beginning of this
section, we have to determine the quantum correlations of the
maximally entangled state of two subsystems having $d$ equally
weighted terms in its Schmidt decomposition. According to the
theorem of Bennett {\em et al.} \cite{Ben96}, the MES has $\log
_{2}d$ ebits of entanglement, where $d$ is the Hilbert space
dimension of the smaller subsystem. Thus,  in our case, the MES of
the subsystem $A$ consisting of $M$ dots and the subsystem $B$ of
$N-M$ dots has
\begin{equation}
E_{{\rm MES}}^{NM}=\log _{2}[\min (M,N-M)+1] \label{N12}
\end{equation}
ebits of entanglement. In particular, the MES in the $N$-dot
system with a single initial excitation has only 1 ebit
independent of $N$. The empty staircase in Fig.  4 and solid lines
in Fig. 5 correspond to $E_{{\rm MES}}^{NM}$. To show a deviation
of a given state from the MES, it is convenient to use the
relative (or scaled) entanglement defined to be
\begin{eqnarray}
e^{NM}_{\max} \equiv \frac{E^{NM}_{\max}}{E^{NM}_{{\rm MES}}} =
\max_t \frac{ E^{NM}(t)}{E^{NM}_{\rm MES}}. \label{N13}
\end{eqnarray}

In the simplest nontrivial case, for $M=1$, the Schmidt
coefficients reduce to
\begin{equation}
P_{1}^{N1}(t)=4{{\frac{(N-1)}{N^{2}}}}\sin
^{2}\left(\frac{N}{2}\kappa t\right) \label{N14}
\end{equation}
and $P_{0}^{N1}(t)=1-P_{1}^{N1}(t)$, which enable a direct
calculation of the entanglement $E^{N1}(t)$ with the help of Eq.
(\ref{N10}). The evolutions of entanglement and the Schmidt
coefficients of $P_{m}^{N1}(t)$ for $m$=0,1, are depicted in Fig.
2. The quantum-dot systems evolve into the MESs at evolution times
that are the roots of the equation
\vspace*{0mm}
\begin{figure}
\includegraphics[width=7cm]{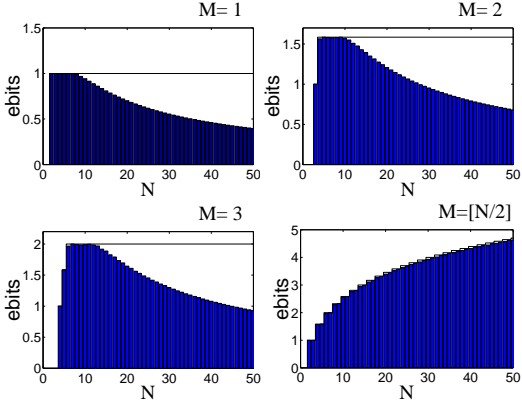}
\caption{Maximum  entanglement $E_{\max }^{NM}$ as a function of
the total number $N$ of dots generated in systems with $M$=1,2,3,
and $[N/2]$ excitations. The solid lines and empty staircase
correspond to $E_{\rm MES}^{NM}$.  On the scale of the figure, an
apparent plateau occurs for $N$ smaller than some critical value
$N_{M}$. For $N$ higher than $N_{M}$ and fixed $M$, a monotonic
decrease of the maximum entanglement is clearly visible. One
concludes that arbitrary high entanglement can be achieved by
increasing $N$ and keeping half $M=[N/2] $ of the system excited.}
\end{figure}
\vspace*{0mm}
\begin{figure}
\includegraphics[width=7cm]{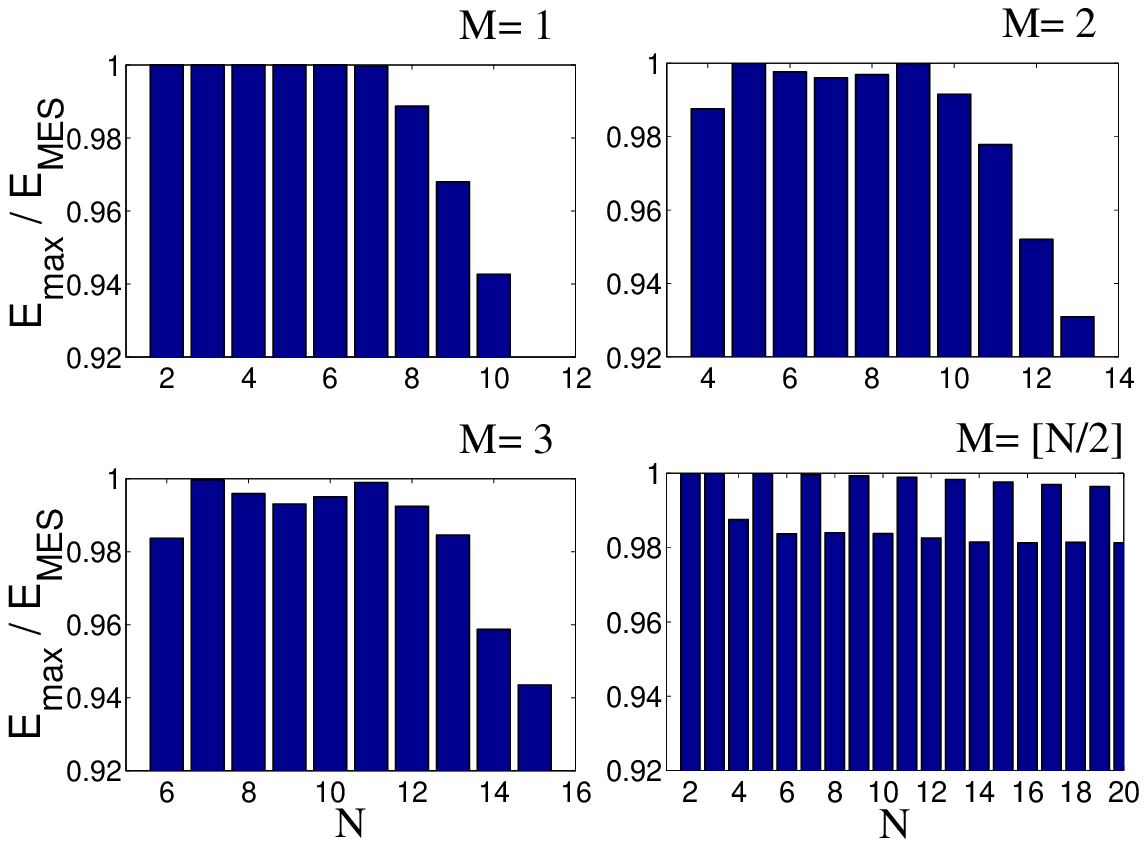}
\caption{The same as in Fig. 5 but for the relative maximum
entanglement $e_{\max }^{NM}=E_{\max }^{NM}/E_{\rm MES}^{NM}$. The
figure shows that the apparent plateau for finite $M$ actually
occurs for $M=1$ only. The first and second highest maxima of
entanglement correspond to $N$ equal to $2M+1$ and $2M+5$,
respectively.}
\end{figure}
\begin{eqnarray}
0=\dot{E}^{NM}(t)&=&2\kappa{{\frac{N-1}{N}}}\sin (N\kappa t)  \nonumber \\
&\times&\!\! \log _{2}\left[ {{\frac{N^{2}}{4(N-1)}}}\csc
^{2}\left({{\frac{N}{2}}}\kappa t\right)-1 \right]. \label{N15}
\end{eqnarray}
Thus, we get
\begin{equation}
\kappa t^{\prime }={{\frac{2}{N}}}{\rm
arccsc}\left({{\frac{2}{N}}}\sqrt{2(N-1)}\right) \label{N16}
\end{equation}
and $\kappa t^{\prime \prime }=\pi/N$. We find that the maximum
entanglement, equal to ${E}^{N1}(t^{\prime })=1$ ebit, can be
achieved at evolution times $t^{\prime}$ for $N\leq 6$ only. For
$N>6$, a real solution for $t^{\prime}$ does not exist. Another
explanation of this result, as illustrated in Fig. 2, can be given
as follows: The maximum entanglement corresponds to the Schmidt
coefficients mutually equal or, in general, the least different.
But the MES corresponds solely to the former case. As seen in Fig.
2, the condition $P_{0}^{N1}(t')= P_{1}^{N1}(t')$ is strictly
satisfied for $N\le 6$. The entanglement for $N>6$ reaches its
maximum at evolution times $t^{\prime \prime }$. This maximum
value is given by
\begin{eqnarray}
{E}^{N1}(t^{\prime \prime }) &=&{{\frac{2}{N^{2}}}}\big\{N^{2}\log
_{2}N-(N-2)^{2}\log _{2}(N-2)  \nonumber \\
&&~~~~~-2(N-1)\log _{2}[4(N-1)]\big\}, \label{N17}
\end{eqnarray}
which is less than unity and monotonically decreases with
increasing $N$ as clearly illustrated in Figs. 5 and 6 for $M$=1.
Thus, the perfect MESs cannot be generated in systems of $N>6$
dots. Nevertheless, a good approximation of the MESs can also be
obtained for $N=7$. On the scale of Fig. 2, $\max_t
E^{7,1}(t)=E^{7,1}(\pi/7)=0.9997$ is close to unity since
$P_{0}^{N1}(\pi/7)$ and $P_{1}^{N1}(\pi/7)$ are almost the same.
It is worth noting that a critical value of $N=6$ was also found,
although in the different context of the pairwise entanglement
measured by the concurrence \cite{Woo98}, for an
equivalent-neighbor model of entangled webs in Ref. \cite{Koa00}.
In comparison, a critical value of $N=6$ for the concurrence in
the equivalent-neighbor isotropic or anisotropic Heisenberg models
was not observed (see, e.g., \cite{Wang01}). Similarly, generation
of the MESs  in an equivalent-neighbor quantum-dot model of Reina
{\em et al.} was discussed only in two special cases of the Bell
($N$=2) and Greenberger-Horne-Zeilinger (GHZ) ($N$=3) entangled
states \cite{Rei00}. Thus, no critical behavior of entanglement as
a function of $N$ was reported there.

The case for $M=1$ is the only one where the general formula
(\ref{N09}) for the Schmidt coefficients simplifies to a compact
form for arbitrary evolution times. Thus, for clarity, we present
mainly numerical results for $M\ge 2$. For example, Fig. 3
illustrates that the exact MESs cannot be generated in systems
with $M=2$ excitations at any evolution time. This conclusion can
be drawn from the observation that $P_{m}^{N2}(t)$ for $m$=0,1,2
do not cross simultaneously at any times in the period.
Nevertheless, the MESs can be approximated with good precision.
The highest possible entanglement, corresponding to the least
mutually different $P_{m}^{N2}$, is observed for $N=5$ and 9,
where the relative entanglement deviates from unity at the order
of $10^{-5}$ and $10^{-4}$, respectively (see Fig. 6 for $M$=2).
The states generated in $N$-dot systems with three excitations can
be entangled up to $e^{7,3}_{\max}=0.9996$ (first) and
$e^{11,3}_{\max}=0.9990$ (second maximum) for the relative
entanglement (see Fig. 6 for $M$=3). It is interesting to compare
the relative entanglement of $e^{NM}_{\max}$, depicted in Fig. 6,
with the ``absolute'' entanglement of $E^{NM}_{\max}$ presented in
Fig. 5. By analyzing the numerical data given, in part, in Fig. 6,
we find the following rule: The maximally or almost maximally
entangled states can be generated in systems of $N=2M+1$ dots with
$M$ excitations. Slightly worse entanglement can be achieved in
systems of $N=2M+5$ dots with $M$ excitations. Thus, systems
composed of odd rather than even numbers of dots enable generation
of the entangled states better approximating the MESs for $M>1$.
This is clearly illustrated in Fig. 6 for $M=[N/2]$, i.e., the
integer part of $N/2$. We observe that the system of odd and large
numbers ($N>2M+5$ for $M>1$) of dots is the most entangled at
evolution times $\kappa t=(1+2k)\pi$ for $k= 0,1,...$ (see, e.g.,
Fig. 3 for $N$=11). In this special case, the Schmidt coefficients
can be written compactly via
\begin{eqnarray}
\left|C_{m}^{NM}\left(\frac{\pi}{\kappa}\right)\right|=2^mm!(N-2M)\frac{(N-2m-2)!!}{N!!}.
\label{N18}
\end{eqnarray}
For $\kappa t=k\pi$ and even $N$, in contrast to odd $N$, the
entanglement vanishes. The maximum entanglement of $E^{NM}_{\max}$
for $N>N_{M}\equiv 2M+5-\delta_{1M}$ can be well fitted by the
inverse of linear functions as shown in Fig. 7.

\vspace*{0mm}
\begin{figure}
\includegraphics[width=8cm]{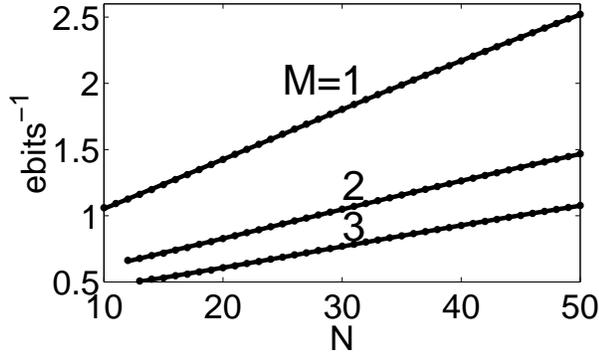}
\caption{The inverse of the maximum entanglement,
$(E_{\max}^{NM})^{-1}$ (dots) measured in ebits$^{-1}$,  and its
approximation (solid lines) as a function of $N>2M+5$ generated in
systems with $M=1,2,3$ excitations.}
\end{figure}

\section{Conclusion}

We studied the evolution of the conduction-band spins of quantum
dots in the model of Imamo\={g}lu {\em et al.} \cite{Ima99}. We
found the analytical solution and its Schmidt decomposition for
the equivalent-neighbor model and applied them in our study of
bipartite entanglement in quantum-dot systems with arbitrary
numbers of dots and their excitations. We have raised and solved
the problem to what extent the initially excited dots can be
entangled with the remaining dots if all of them are initially
disentangled in the equivalent-neighbor energy-conserving model.
We have shown that the perfect maximally entangled states can only
be generated in systems of $N=2,\cdots ,6$ dots with a single dot
initially excited. Nevertheless, highly entangled states, being
excellent approximations of the MESs, can periodically be
generated in systems of odd numbers $N$ of dots with the number
$M$ of excitations equal to $M=(N-1)/2$ (leading to the best
approximation) and $M=(N-5)/2$ (giving a slightly worse
approximation). If we increase $N$ beyond
$N_{M}=2M+5-\delta_{1M}$, the entanglement decreases monotonically
as described by the inverse of linear functions.

\vspace*{10mm}
\section*{ACKNOWLEDGMENTS}
We thank J. Bajer, T. Cheon, T. Kobayashi, H. Matsueda, and I.
Tsutsui for their stimulating discussions. Y.L. acknowledges
support from the Japan Society for the Promotion of Science
(JSPS). This work was supported by a Grant-in-Aid for Scientific
Research (B) (Grant No.~12440111) and a Grant-in-Aid for
Encouragement of Young Scientists (Grant No. 12740243) by the
Japan Society for the Promotion of Science.

\vspace*{0mm}


\widetext
\vspace{1mm} {\setlength{\fboxsep}{1pt}
\begin{center}
\framebox{\parbox{0.75\columnwidth}{%
\begin{center}
published in the {\em Physical Review A} {\bf 65} (2002) 062321\\
and selected to\\
{\em Virtual J. Nanoscale Sci. Tech.}
({\tt http://www.vjnano.org/nano/}) {\bf 6} (2002) Issue 1; \\
{\em  Virtual J. Quantum Information} {\tt
(http://www.vjquantuminfo.org)} {\bf 2} (2002) Issue 7.
\end{center}}}
\end{center}

\end{document}